\documentclass[12pt]{article}
\title{Universal Probability Distribution for the Wave Function of a Quantum
System Entangled with Its Environment}
\author{
Sheldon Goldstein\footnote{Departments of Mathematics and Physics,
     Rutgers University, Hill Center, 
     110 Frelinghuysen Road, Piscataway, NJ 08854-8019, USA.
     E-mail: oldstein@math.rutgers.edu},
Joel L. Lebowitz\footnote{Departments of Mathematics and Physics,
     Rutgers University, Hill Center, 
     110 Frelinghuysen Road, Piscataway, NJ 08854-8019, USA.
     E-mail: lebowitz@math.rutgers.edu},
Christian Mastrodonato\footnote{Dipartimento di Fisica, Universit\`a di
    Genova and INFN sezione di Genova, Via Dodecaneso 33, 16146
    Genova, Italy.  E-mail: christian.mastrodonato@ge.infn.it},\\
Roderich Tumulka\footnote{Department of Mathematics,
     Rutgers University, Hill Center, 
     110 Frelinghuysen Road, Piscataway, NJ 08854-8019, USA.
     E-mail: tumulka@math.rutgers.edu}, and
Nino Zangh\`\i\footnote{Dipartimento di Fisica, Universit\`a di
    Genova and INFN sezione di Genova, Via Dodecaneso 33, 16146
    Genova, Italy.  E-mail: zanghi@ge.infn.it}
}
\date{February 10, 2015}

\usepackage{amsthm,url,amssymb,amsmath,amsfonts,mathrsfs}
\usepackage[usenames,dvipsnames]{color}

\theoremstyle{plain}
\newtheorem{thm}{Theorem}
\newtheorem{lemma}{Lemma}

\newcommand{\ket}[1]{\vert#1\rangle}
\newcommand{\bra}[1]{\langle#1\vert}
\DeclareMathOperator{\tr}{tr}
\DeclareMathOperator{\Var}{Var}
\DeclareMathOperator{\Cov}{Cov}
\newcommand{\CCC}{\mathbb{C}}
\newcommand{\RRR}{\mathbb{R}}
\newcommand{\NNN}{\mathbb{N}}
\newcommand{\EEE}{\mathbb{E}}
\newcommand{\PPP}{\mathbb{P}}
\renewcommand{\Re}{\mathrm{Re}}

\newcommand{\scp}[2]{\langle #1| #2 \rangle}
\newcommand{\Hilbert}{\mathscr{H}}
\newcommand{\total}{\mathrm{total}}
\newcommand{\sphere}{\mathbb{S}}

\newcommand{\D}{\mathscr{D}}
\newcommand{\G}[1]{G(#1)}
\newcommand{\GA}[1]{GA(#1)}
\newcommand{\GAP}[1]{GAP(#1)}
\newcommand{\be}{\begin{equation}}
\newcommand{\ee}{\end{equation}}

\newcommand{\y}[1]{{#1}}

\begin{document}
\maketitle
\begin{abstract}
A quantum system (with Hilbert space $\Hilbert_1$) entangled with its environment (with Hilbert space $\Hilbert_2$) is usually not attributed a wave function but only a reduced density matrix $\rho_1$. Nevertheless, there is a precise way of attributing to it a random wave function $\psi_1$, called its conditional wave function, whose probability distribution $\mu_1$ depends on the entangled wave function $\psi\in\Hilbert_1\otimes\Hilbert_2$ in the Hilbert space of system and environment together. It also depends on a choice of orthonormal basis of $\Hilbert_2$ but in relevant cases, as we show, not very much. We prove several universality (or typicality) results about $\mu_1$, e.g., that if the environment is sufficiently large then for every orthonormal basis of $\Hilbert_2$, most entangled states $\psi$ with given reduced density matrix $\rho_1$ are such that $\mu_1$ is close to one of the so-called GAP (Gaussian adjusted projected) measures, $GAP(\rho_1)$. We also show that, for most entangled states $\psi$ from a microcanonical subspace (spanned by the eigenvectors of the Hamiltonian with energies in a narrow interval $[E,E+\delta E]$) and most orthonormal bases of $\Hilbert_2$, $\mu_1$ is close to $GAP(\tr_2 \rho_{mc})$ with $\rho_{mc}$ the normalized projection to the microcanonical subspace. In particular, if the coupling between the system and the environment is weak, then $\mu_1$ is close to $GAP(\rho_\beta)$ with $\rho_\beta$ the canonical density matrix on $\Hilbert_1$ at inverse temperature $\beta=\beta(E)$. This provides the mathematical justification of our claim in \cite{Gold1} that $GAP$ measures describe the thermal equilibrium distribution of the wave function.

\medskip

\end{abstract}

\tableofcontents

\section{Introduction}

In this paper we establish the \emph{universality} of certain probability distributions on Hilbert spaces known as \emph{Scrooge measures} or \emph{Gaussian adjusted projected (GAP) measures} \cite{JRW94, Gold1,Rei08} (see Section~\ref{sec:GAPdef} below for the definition). This makes precise some statements and mathematical considerations discussed in our earlier paper \cite{Gold1}; our main physical conclusion, elucidated in Section~\ref{sec:summary} below, is that the wave function of an open quantum system (i.e., a subsystem of a larger system) possesses a thermal equilibrium distribution given by a GAP measure.

By the wave function of a subsystem, we mean more precisely the \emph{conditional wave function}, described in Section~\ref{sec:condwf} below. 
By saying that GAP measures are universal we mean that, when the system's environment is sufficiently large, the distribution $\mu_1$ of the conditional wave function is typically close to a GAP measure, namely
\be
\mu_1 \approx GAP(\rho_1)
\ee
(see below). To illustrate the terminology of universality, one can say that the central limit theorem conveys a sense in which the Gaussian probability distribution on the real line is universal: many physically relevant probability distributions are approximately Gaussian. Instead of universality, one also often speaks of typicality; we use these two terms more or less interchangeably.

The family of GAP measures is a family of probability measures on Hilbert spaces. There is one GAP measure for every density matrix $\rho$ on a Hilbert space $\Hilbert$, denoted $GAP(\rho)$; it is concentrated on the unit sphere in $\Hilbert$,
\begin{equation}
  \sphere(\Hilbert) = \{\psi \in \Hilbert: \|\psi\|=1\}\,.
\end{equation}
The density matrix of $GAP(\rho)$ is $\rho$. By this we mean the following: For any probability measure $\mu$ on $\sphere(\Hilbert)$, its density matrix is
\begin{equation}\label{covmatrix}
\rho_\mu=\int_{\sphere(\Hilbert)} \mu(d\psi) \ket{\psi}\bra{\psi}\,,
\end{equation}
which is also the covariance matrix of $\mu$ provided $\mu$ has mean zero. For $\mu=GAP(\rho)$, $\rho_\mu=\rho$.

Of particular interest are the GAP measures associated with canonical density matrices
\begin{equation}\label{rhobeta}
  \rho_\beta = \frac{1}{Z} e^{-\beta H}\,,
\end{equation}
where $Z= \tr e^{-\beta H}$ is the normalization constant,
$\beta$ the inverse temperature and $H$ the Hamiltonian. 
Our main conclusion is that the conditional wave function of a system entangled with its environment is $GAP(\rho_\beta)$-distributed for pure states of the system and environment that correspond to thermal equilibrium. Detailed
discussions of GAP measures and their physical applications can be 
found in \cite{Gold1,Rei08}. See
\cite{TZ05} for a study about the support of GAP measures, that is,
about what $GAP(\rho_\beta)$-distributed wave functions typically look like.

The main application of GAP measures
is the characterization of the wave functions of systems we encounter in
nature. In most cases we do not know a system's wave function,
but in many cases the system is more or less in thermal equilibrium, and then, according to the considerations presented in \cite{Gold1} and here,
its wave function should be GAP distributed.

\subsection{Conditional Wave Function}
\label{sec:condwf}

Consider a composite quantum system consisting of two subsystems, system 1 and system 2, with associated Hilbert spaces $\Hilbert_1$ and $\Hilbert_2$. Suppose that the system is in a pure state $\psi \in \Hilbert_{\total} = \Hilbert_1 \otimes \Hilbert_2$. We ask what might be meant by the wave
function of system 1. An answer is provided by the notion of conditional wave function, defined as follows \cite{GN99,Gold1}:\footnote{This definition is inspired by Bohmian mechanics, a formulation of quantum mechanics with particle trajectories, where the (non-normalized) conditional wave function $\psi_1$ of system 1 is defined by \cite{DGZ}
\begin{equation}
  \psi_1(x) = \psi(x,Y)
\end{equation}
for $x$ in the configuration space of system 1, with $Y$ the actual configuration of system 2. If system 2 contains particles with spin, then the configuration basis is not a basis of $\Hilbert_2$, and we may either choose a basis of spin space or trace out the spin indices; see \cite{PT13} for a discussion of the latter choice.} 
Let $b=\{b_j\}$ be an orthonormal basis of $\Hilbert_2$. For
each choice of $j$, the partial inner product $\scp{b_j}
{\psi}$, taken in $\Hilbert_2$, is a vector belonging to
$\Hilbert_1$. Regarding $j$ as random (and therefore writing $J$), we are led to
consider the random vector $\psi_1\in \Hilbert_1$ given by
\begin{equation}\label{Psi1def}
  \psi_1 = \frac{\scp{b_J} {\psi}}{\bigl\|\scp{b_J} {\psi}\bigr\|}
\end{equation}
where 
$b_J$ is a random element of the basis $\{b_j\}$, chosen with the quantum distribution
\begin{equation}\label{marg}
  \PPP^{\psi,b}(J = j) = \bigl\| \scp{b_j} {\psi} \bigr\|^2.
\end{equation}
We refer to $\psi_1$ as the conditional wave function of system 1.\footnote{The conditional wave function can be regarded as a precise version of the ``collapsed'' wave function in the standard quantum formalism: Suppose that system 1 has interacted with system 2, and their joint wave function, as produced by the appropriate Schr\"odinger evolution, is now 
\be\label{superposition}
\sum_j c_j \psi_j^{(1)} \otimes \psi_j^{(2)}\,,
\ee
where the $c_j$ are complex coefficients and all $\psi$s are normalized. If system 2 is a macroscopic system and the $\psi_j^{(2)}$s are macroscopically different states then in the standard formalism one regards $j$ as random with distribution $|c_j|^2$, and says accordingly that system 1 can be attributed the ``collapsed'' wave function $\psi_j^{(1)}$ with probability $|c_j|^2$. The conditional wave function of system 1, according to the above definition in the case that the $\psi_j^{(2)}$s are among the $\{b_j\}$, is indeed $\psi_j^{(1)}$ with probability $|c_j|^2$.}

The distribution of $\psi_1$ corresponding to \eqref{Psi1def} and
\eqref{marg} is given by the following probability measure on $\sphere
(\Hilbert_1)$: The probability that $\psi_1\in A\subseteq \sphere(\Hilbert_1)$ is
\begin{align}\label{mu1}
  \mu_1(A)=\mu_1^{\psi,b}(A)= \PPP(\psi_1 \in A) 
  &= \sum_{j} \bigl\|
  \scp{b_j} {\psi} \bigr\|^2 \, \delta_{\scp{b_j}{\psi}/\|\scp{b_j}{\psi}\|}(A) \\
  &= \sum_{j} \bigl\|
  \scp{b_j} {\psi} \bigr\|^2 \, 1_A\biggl(\frac{\scp{b_j}{\psi}}{\|\scp{b_j}{\psi}\|}\biggr) \,,\label{mu1expr}
\end{align}
where $\delta_\phi$ denotes the Dirac ``delta'' measure (a point
mass) concentrated at $\phi$ and $1_A$ denotes the characteristic function of the set $A$. While the density matrix $\rho_{\mu_1}$ associated with ${\mu_1}$ always equals the reduced
density matrix $\rho_1^\psi$ of system 1, given by
\begin{equation}\label{rhoreddef}
  \rho_1^\psi = \tr_2 |\psi \rangle \langle \psi | = \sum_{j}
  \scp{b_j}{\psi} \scp{\psi}{b_j} \,,
\end{equation}
the measure $\mu_1$ itself usually depends on the choice of the basis $b$, so $\mu_1=\mu_1^{\psi,b}$.

\subsection{Summary of Results}
\label{sec:summary}

In this paper, we prove several \emph{universality theorems} about GAP measures, Theorems~\ref{thm1}--\ref{thm4}, formulated in Section~\ref{sec:results}. These are statements to the effect that for \emph{most} wave functions $\psi$ from relevant subsets of $\Hilbert_1\otimes\Hilbert_2$ and/or \emph{most} orthonormal bases $b$ of $\Hilbert_2$, $\psi_1$ is approximately GAP-distributed. Here, ``most'' means that the set of exceptions is small with respect to the appropriate natural uniform measure. 

The basic universality property is expressed in Theorem~\ref{thm1}, which asserts that for sufficiently large $\dim \Hilbert_2$, for any orthonormal basis $b$ of $\Hilbert_2$, and for any density matrix $\rho_1$ on $\Hilbert_1$, most $\psi$ in $\sphere(\Hilbert_1\otimes\Hilbert_2)$ with the reduced density matrix $\tr_2\ket{\psi}\bra{\psi} = \rho_1$ are such that the distribution $\mu_1^{\psi,b}$ of $\psi_1$ is arbitrarily close to $GAP(\rho_1)$,
\be\label{mu1GAPrho1}
\mu_1^{\psi,b} \approx GAP(\rho_1)\,.
\ee
This fact was derived (but not rigorously proven) in Section 5.1.3 of \cite{Gold1}.

Theorem~\ref{corbasis} asserts that the conclusion of Theorem~\ref{thm1}---that \eqref{mu1GAPrho1} holds with arbitrary accuracy for sufficiently large $\dim\Hilbert_2$---is also true for \emph{every} $\psi$ with $\tr_2 \ket{\psi}\bra{\psi}=\rho_1$ for \emph{most} $b$ (instead of for \emph{every} $b$ for \emph{most} $\psi$).

Theorems~\ref{thm3} and \ref{thm4} justify the conclusion that, if a system (system 1) is weakly coupled to a very large (but finite) second system
then, for most wave functions of the composite system with energy in a given narrow energy range $[E,E+\delta E]$, the conditional wave function of the system is approximately GAP-distributed for most orthonormal bases of system 2. In more detail, let the interaction between the two systems be negligible so that the Hamiltonian can be taken to be
\be\label{noint}
H=H_1\otimes I_2 + I_1\otimes H_2
\ee
(with $I_{1/2}$ the identity operator on $\Hilbert_{1/2}$), and let $\Hilbert_R\subset \Hilbert_1\otimes \Hilbert_2$ be a micro-canonical energy shell of the composite system, i.e., the subspace spanned by the eigenstates of the total energy with eigenvalues in $[E,E+\delta E]$. Assume that the eigenvalues of $H_2$ are sufficiently dense and that the dimensions of $\Hilbert_2$ and $\Hilbert_R$ are sufficiently large. Then, for most $\psi\in\sphere(\Hilbert_R)$,
\be
\mu_1^{\psi,b} \approx GAP(\rho_\beta)
\ee
for most bases $b$ of $\Hilbert_2$; here, $\rho_\beta$ is the canonical density matrix \eqref{rhobeta} and $\beta=\beta(E)$ . 

In Theorems~\ref{thm3} and \ref{thm4} we relax the condition that $\psi$ have a prescribed reduced density matrix, and exploit instead \emph{canonical typicality}. This is the fact, found independently by several groups \cite{GMM04,Gold2,PSW05,PSW06}
and anticipated long before by Schr\"odinger \cite{schrbook},
that for most $\psi\in\sphere(\Hilbert_R)$, the reduced density matrix $\tr_2\ket{\psi}\bra{\psi}$ is approximately of the canonical form \eqref{rhobeta}.
More generally, in Theorems~\ref{thm3} and \ref{thm4} we may regard $\Hilbert_R$ as \emph{any} subspace of $\Hilbert_1\otimes\Hilbert_2$ of sufficiently high dimension. Canonical typicality then refers to the fact that for most $\psi\in\sphere(\Hilbert_R)$,
$\tr_2\ket{\psi}\bra{\psi}$ is close to $\tr_2\rho_R$, where $\rho_R$ denotes $1/\dim\Hilbert_R$ times the projection to $\Hilbert_R$; the precise version of canonical typicality that we use in the
proof of our Theorems~\ref{thm3} and \ref{thm4} is due to Popescu, Short,
and Winter \cite{PSW05,PSW06}. 

Theorem~\ref{thm3} asserts that for most $\psi\in\sphere(\Hilbert_R)$,
\be\label{mu1GAPtr2rhoR}
\mu_1^{\psi,b} \approx GAP(\rho^{(1)}_R)\,,
\ee
with $\rho^{(1)}_R=\tr_2 \rho_R$, for most orthonormal bases $b$ of $\Hilbert_2$. This means in particular that we need not restrict ourselves to weak (relatively negligible) interactions between systems 1 and 2 as in \eqref{noint}.

Theorem~\ref{thm4} is a very similar statement but differs in the detailed meaning of ``$\approx$'' and refers to a fixed density matrix, such as $\rho_\beta$, in place of $\rho^{(1)}_R$ in \eqref{mu1GAPtr2rhoR}.

\y{Theorems 3 and 4 follow from Theorem 2 by means of canonical typicality and continuity of the mapping $\rho\mapsto GAP(\rho)$. However, we need to pay careful attention here to the details, in particular to the various possible meanings of ``continuity,'' corresponding to various topologies over measures, involving various classes of test functions, uniformity in $\rho$ or in the test function, and domains of that uniformity, for example.}

\subsection{Remarks}

\begin{itemize}
\item \textit{Time evolution.} It may be interesting to consider how $\mu_1^{\psi,b}$ evolves with time if the wave function $\psi=\psi_t$ of systems 1 and 2 together evolves according to the Schr\"odinger equation
\be\label{Schr}
i\hbar\frac{\partial \psi_t}{\partial t} = H \psi_t\,.
\ee
In a situation in which \y{$\Hilbert_R$ is a micro-canonical energy shell (and thus invariant under \eqref{Schr}), and} most $\psi\in\sphere(\Hilbert_R)$ have $\mu_1^{\psi,b}\approx GAP(\rho^{(1)}_R)$, we may expect that even for $\psi_0\in\sphere(\Hilbert_R)$ with $\mu_1^{\psi_0,b}$ far from any GAP measure, $\mu_1(t)=\mu_1^{\psi_t,b}$ will approach $GAP(\rho^{(1)}_R)$ and stay near $GAP(\rho^{(1)}_R)$ most of the time (though not forever, as follows from the recurrence property (almost-periodicity) of the Schr\"odinger evolution in a finite-dimensional Hilbert space). We leave this problem open but briefly remark that one can already conclude by interchanging the time average and the average over $\psi_0$ that whenever it is true for most $\psi\in\sphere(\Hilbert_R)$ that $\mu_1^{\psi,b}\approx GAP(\rho^{(1)}_R)$, then for most $\psi_0\in\sphere(\Hilbert_R)$, $\mu_1^{\psi_t,b} \approx GAP(\rho^{(1)}_R)$ for most times $t$; the open problem is to prove a statement that concerns \emph{all}, rather than \emph{most}, $\psi_0$ (under suitable hypotheses).

\item \textit{The role of interaction.}
Another remark concerns the role of interaction (between the system and its environment) for obtaining the distribution $GAP(\rho_\beta)$. The nature of the interaction is relevant to our discussion in two places---although our theorems do not depend on it, as they do not mention the Hamiltonian at all. First, interaction is relevant for creating typical wave functions, as it helps evolve atypical wave functions into typical ones. This is closely related to the fact that a system coupled to a big second system will typically go from non-equilibrium to thermal equilibrium only in the presence of interaction; see Section~4 of \cite{GLMTZ09b} for further discussion and examples. Second, it depends on the interaction which subspace of $\Hilbert_1\otimes\Hilbert_2$ is the micro-canonical energy shell that we want $\Hilbert_R$ to be, and thus also which density matrix $\tr_2 \rho_R$ is. In the limit of negligible interaction, $\tr_2 \rho_R$ has the canonical form $\rho_\beta = (1/Z) e^{-\beta H}$, while interaction makes it deviate from this form. As a consequence of these two roles, when we want to obtain from non-equilibrium a wave function $\psi\in\Hilbert_1\otimes\Hilbert_2$ such that the distribution of the conditional wave function $\psi_1$ is close to $GAP(\rho_\beta)$, we may want that the interaction be not too large (or else there will be deviations from $\rho_\beta$) and that the interaction be not too small (or else it may take too long, say longer than the present age of the universe, to reach thermal equilibrium). 

\end{itemize}

\subsection{Definition of the GAP Measure}
\label{sec:GAPdef}

Let $\Hilbert$ be a Hilbert space and $\rho$ a density matrix on $\Hilbert$. We describe four equivalent definitions of the measure $GAP(\rho)$ on (the Borel $\sigma$-algebra of) $\sphere(\Hilbert)$.

The first definition involves Gaussian measures and proceeds in three steps represented by the acronym $GAP$. We start from the measure $G(\rho)$,
which is the Gaussian measure on $\Hilbert$ with mean 0 and covariance matrix
$\rho$. In this paper, we are interested only in the case $\dim\Hilbert< \infty$.
Then $G(\rho)$ can be explicitly defined as follows: Let $S$ be the subspace of 
$\Hilbert$ on which $\rho$ is supported, i.e., its positive spectral subspace, or
equivalently the orthogonal complement of its kernel, or
equivalently its range; let $d'=\dim S$ and $\rho_+$ the restriction
of $\rho$ to $S$; then $G(\rho)$ is the measure on $\Hilbert$
supported on $S$ with the following density relative to the Lebesgue
measure $\lambda$ on $S$:
\begin{equation}
  \frac{dG(\rho)}{d\lambda}(\psi) = \frac{1}{\pi^{d'} \,
  \det \rho_+} \exp(-\langle \psi |\rho^{-1}_+| \psi \rangle)\,.
\end{equation}
Equivalently, a $G(\rho)$-distributed random vector $\psi$ is one whose coefficients $\scp{\chi_i}{\psi}$ relative to an eigenbasis $\{\chi_i\}$ of $\rho$ (i.e., $\rho\chi_i = p_i \chi_i$ with $0\leq p_i\leq 1$) are independent complex Gaussian random variables with mean 0 and variances $\EEE|\scp{\chi_i}{\psi}|^2=p_i$; by a \emph{complex Gaussian} random variable we mean one whose real and imaginary parts are independent real Gaussian random variables with equal variances.

Noting that 
\be\label{Gpsi1}
\int_\Hilbert G(\rho)(d\psi) \, \|\psi\|^2 = \tr \rho =1\,,
\ee
we now define the adjusted Gaussian measure $GA(\rho)$ on $\Hilbert$ as:
\begin{equation}
  GA(\rho)(d\psi) = \|\psi\|^2 G(\rho)(d\psi)\,.
\end{equation}
If $\psi^{GA}$ is a $GA(\rho)$-distributed vector, then $GAP(\rho)$
is the distribution of this vector projected on the unit sphere; that is, $GAP(\rho)$ is the distribution of
\begin{equation}
  \psi^{GAP} = \frac{\psi^{GA}}{\|\psi^{GA}\|}\,.
\end{equation}
Like $\G{\rho}$ and unlike $\GA{\rho}$, $\GAP{\rho}$ has covariance matrix $\rho$.

More generally, one can define for any measure $\mu$ on $\Hilbert$
the ``adjust-and-project'' procedure, \y{producing a measure we sometimes denote by $\mu AP$.} We denote by $A\mu$ the
adjusted measure
\begin{equation}\label{Adef}
A\mu(d\psi) = \|\psi\|^2 \, \mu(d\psi)\,.
\end{equation}
The projection on the unit sphere is defined as:
\begin{equation}\label{Pdef}
P:\Hilbert \setminus \{0\} \to \sphere(\Hilbert)\, , \quad
P(\psi)= \frac{\psi}{\|\psi\|}\,.
\end{equation}
Then the adjusted-and-projected measure is $\mu AP=P_* ( A\mu) = A\mu \circ
P^{-1}$, where $P_*$ denotes the action of $P$ on measures, thus
defining a mapping $P_* \circ A$ from the
measures on $\Hilbert$ 
to the measures on $\sphere(\Hilbert)$.
If $\int \mu(d\psi) \, \|\psi\|^2 = 1$ then $P_*(A\mu)$ is a probability measure.

The second definition \cite{JRW94} works without Gaussian measures; it applies when $d:=\dim\Hilbert<\infty$. Let $\Psi^u$ be uniformly distributed on $\sphere(\Hilbert)$, and let $uD(\rho)$ denote the distribution of
\be
\Psi^{uD(\rho)} = d^{1/2}\rho^{1/2} \Psi^u\,.
\ee
It is a measure on $\Hilbert$ concentrated on the ellipsoid that is the image of the unit sphere under $d^{1/2}\rho^{1/2}$. Then (as shown below)
\be\label{uDAP=GAP}
uDAP(\rho)=GAP(\rho)\,.
\ee
That is, applying the adjust-and-project procedure to $uD(\rho)$ yields $GAP(\rho)$. 

More generally, Jozsa et al.~\cite{JRW94} defined for any probability measure $\mu$ on $\sphere(\Hilbert)$ the procedure of \emph{$\rho$-distortion}, yielding $\mu DAP(\rho)$, as follows: Let $\Psi^\mu$ be $\mu$-distributed and let $\mu D(\rho)$ denote the distribution of
\be
\Psi^{\mu D(\rho)} = d^{1/2} \rho^{1/2} \Psi^\mu\,.
\ee
Then apply the adjust-and-project procedure to obtain the measure $\mu DAP(\rho)$ (which in general is not normalized) on $\sphere(\Hilbert)$. In these terms, $GAP(\rho)$ is the $\rho$-distortion of the uniform probability measure.

The third definition of $GAP(\rho)$ was suggested to us by an anonymous referee; like the previous one, it applies if $d=\dim\Hilbert<\infty$. Let $\Hilbert_2$ be a Hilbert space of the same dimension $d$, and fix a vector $\Phi\in\sphere(\Hilbert\otimes\Hilbert_2)$ such that $\tr_2 |\Phi\rangle\langle\Phi|=\rho$. Choose a random $\Psi_2\in\sphere(\Hilbert_2)$ with distribution 
\be\label{Psi2def}
\mu_2(d\psi_2) = d\,  \bigl\|\scp{\psi_2}{\Phi}\bigr\|^2\, u_2(d\psi_2)\,,
\ee
where $\scp{\cdot}{\cdot}$ is the partial inner product in $\Hilbert_2$, $\|\cdot\|$ is the norm in $\Hilbert$, and $u_2$ is the uniform probability distribution on $\sphere(\Hilbert_2)$; $\mu_2$ is normalized because
\be
\mu_2(\sphere(\Hilbert_2)) = d\int\limits_{\sphere(\Hilbert_2)} \!\!\! u_2(d\psi_2) \, \scp{\Phi}{\psi_2} \scp{\psi_2}{\Phi}
= d \,\scp{\Phi}{d^{-1}I_2|\Phi}=1\,.
\ee
Then $GAP(\rho)$ is the distribution of
\be\label{GAP3Psi}
\Psi= \frac{\scp{\Psi_2}{\Phi}}{\bigl\| \scp{\Psi_2}{\Phi} \bigr\|}\,,
\ee
where $\scp{\cdot}{\cdot}$ is again the partial inner product in $\Hilbert_2$. 
We will prove the equivalence of this definition with the previous one in the next section. 

The measure $\mu_2$ possesses the following operational interpretation. For any Hilbert space $\Hilbert$ with finite dimension $d$, let $E_\Hilbert$ be the unique unitary-covariant positive-operator-valued measure (POVM) on $\sphere(\Hilbert)$ acting on $\Hilbert$; it is defined by
\be
E_\Hilbert(d\psi) = d\, \ket{\psi} \bra{\psi}\, u(d\psi)\,,
\ee
where $u$ is the uniform probability measure on $\sphere(\Hilbert)$. In our setting involving $\Phi\in\sphere(\Hilbert\otimes\Hilbert_2)$, $\mu_2$ coincides with the distribution of the outcome of a (generalized) quantum measurement of $I_1\otimes E_{\Hilbert_2}$ on a system with pure state $\Phi$; the collapsed state after the measurement is then $\Psi\otimes \Psi_2$ as in \eqref{GAP3Psi}.

The fourth definition requires that $d=\dim\Hilbert<\infty$ and that zero is not among the eigenvalues of $\rho$. Then $GAP(\rho)$ possesses a density relative to the uniform probability distribution $u$ on ${\sphere(\Hilbert)}$, which is \cite{Gold1}
\be\label{mupowerlaw}
  \frac{dGAP(\rho)}{du}(\psi) 
  = \frac{d}{\det \rho} \, \langle \psi | \rho^{-1} | \psi \rangle^{-d-1}\,.
\ee

\subsection{Properties of the GAP Measure}
\label{sec:propertiesGAP}

The density matrix associated with $GAP(\rho)$ is $\rho$,
\be
\rho_{GAP(\rho)}=\rho\,.
\ee
To see this, note that the density matrix $\rho_\mu$ as in \eqref{covmatrix} is a special case of the covariance matrix provided $\mu$ has mean 0, and that the covariance matrix can be defined also for probability measures $\mu$ on $\Hilbert$ with mean 0 by
\be
C_\mu =\int\limits_{\Hilbert} \mu(d\psi) \, |\psi\rangle \langle\psi| \,.
\ee
The adjust-and-project procedure preserves the covariance matrix,
\be
C_{P_*(A\mu)}=C_\mu\,,
\ee
for the simple reason \cite{Gold1} that
\be
C_{P_*(A\mu)} 
= \int\limits_{\Hilbert} A\mu(d\psi) \, |P(\psi)\rangle \langle P(\psi)|
= \int\limits_{\Hilbert} \|\psi\|^2\, \mu(d\psi) \, \frac{|\psi\rangle \langle\psi|}{\|\psi\|^2} 
=\int\limits_{\Hilbert} \mu(d\psi) \, |\psi\rangle \langle\psi| =C_\mu \,.
\ee
As a consequence, $\rho_{GAP(\rho)}=C_{GAP(\rho)}=C_{G(\rho)}=\rho$.

If $\rho$ is proportional to a projection, $\rho=(\dim W)^{-1}\, P_W$ for some subspace $W\subseteq \Hilbert$, then $GAP(\rho)=u_{\sphere(W)}$. In general, in a certain precise sense, $GAP(\rho)$ is the most spread-out distribution on $\sphere(\Hilbert)$ with density matrix $\rho$ \cite{JRW94}. Furthermore, the mapping $\rho\mapsto GAP(\rho)$ is covariant under unitary transformations $U$, i.e., $U\Psi^{GAP(\rho)}$ has distribution $GAP(U\rho U^{-1})$ \cite{Gold1}. 

It follows also that the second definition given above is equivalent to the first: 
Note first that $uDAP(\rho)$ is a probability measure because
\be
\EEE\|\Psi^{uD(\rho)}\|^2 
= d\, \EEE \scp{\Psi^u}{\rho|\Psi^u} 
= d\,\EEE\tr\Bigl(\rho\, |\Psi^u\rangle\langle\Psi^u|\Bigr) 
= d \tr(\rho\,  d^{-1}I) 
= \tr \rho =1\,.
\ee
Note also that $\Psi^{G(I/d)}= \Lambda \Psi^u$, where $\Lambda=\|\Psi^{G(I/d)}\|$ is a real-valued random variable independent of $\Psi^u$ with $\EEE\Lambda^2=1$. 
Furthermore, $G(\rho)$ is the distribution of $d^{1/2}\rho^{1/2}\, \Psi^{G(I/d)}=\Lambda d^{1/2}\rho^{1/2}\Psi^u$. The adjustment factor $f(\psi)$ can be written as 
$\Lambda^2 \, d\,\|\rho^{1/2}\Psi^u\|^2$, 
so that $GA(\rho)$ is the distribution of $\tilde\Lambda\, \Psi^{uDA(\rho)}$, where \y{$\Psi^{uDA(\rho)}$ has distribution $uDA(\rho)=A(uD(\rho))$ and} $\tilde\Lambda$ is independent of $\Psi^{uDA(\rho)}$ with $\PPP(\tilde\Lambda\in d\lambda)=\lambda^2\,\PPP(\Lambda\in d\lambda)$. When projecting to $\sphere(\Hilbert)$, the factor $\tilde\Lambda$ cancels out, so that $uDAP(\rho)=GAP(\rho)$.

To see that the third definition is equivalent to the other two, note that $\Phi$ defines an (anti-linear) mapping $\Hilbert_2\to\Hilbert$ by $|\psi_2\rangle \mapsto \scp{\psi_2}{\Phi}$. To express this mapping explicitly using the Schmidt decomposition \cite{Schmidt} of $\Phi$,
\be
\Phi = \sum_i \sqrt{p_i}\, \chi_i\otimes \phi_i
\ee
for some orthonormal basis $\{\phi_i:i=1\ldots d\}$ of $\Hilbert_2$, the vector $\psi_2=\sum_i c_i \,\phi_i$ gets mapped to $\sum_i c_i^* \sqrt{p_i}\, \chi_i$. Put differently, except for the conjugation, the mapping acts like $\rho^{1/2}$. Thus, it maps the distribution $u_2$ to $uD(\rho)$ and $\mu_2$ to $uDA(\rho)$, except for a rescaling in $\Hilbert$ by a factor $d^{1/2}$. The remaining step is the usual projection to $\sphere(\Hilbert)$, which also cancels the $d^{1/2}$.

The last property, expressed by the following lemma, provides a link between the distribution $\mu_1^{\psi,b}$ of the conditional wave function and the $GAP$ measures; it asserts that when $\mu_1^{\psi,b}$ gets averaged over all orthonormal bases $b$ of $\Hilbert_2$, the resulting distribution on $\sphere(\Hilbert_1)$ is a $GAP$ distribution. Let $ONB(\Hilbert_2)$ be the set of orthonormal bases of $\Hilbert_2$, and let $u_{ONB}$ be the uniform probability measure on $ONB(\Hilbert_2)$,
corresponding to the Haar measure on the unitary group $U(\Hilbert_2)$.
Fix $\Hilbert_1,\Hilbert_2$, and $\psi$, and regard $\mu_1^{\psi,b}$ as a function of $b\in ONB(\Hilbert_2)$.

\begin{lemma}\label{lem:EmuGAP}
For $b\sim u_{ONB}$,
\be\label{EmuGAP}
\EEE\, \mu_1^{\psi,b} = \GAP{\rho_1^\psi}\,.
\ee
\end{lemma}

\proof
For any measurable set $A\subseteq \sphere(\Hilbert_1)$, we obtain from the expression \eqref{mu1expr} for $\mu_1^{\psi,b}(A)$, using that the $b_1,\ldots,b_{d_2}$ are exchangeable random vectors, that
\begin{align}
\EEE\,\mu_1^{\psi,b}(A) 
&= d_2 \,\EEE_{b_1\sim u_{\sphere(\Hilbert_2)}} \biggl[ \bigl\| \scp{b_1}{\psi} \bigr\|^2 \, 1_A \biggl( \frac{\scp{b_1}{\psi}}{\|\scp{b_1}{\psi}\|} \biggr) \biggr]\\
&= \EEE_{b_1\sim d_2 \|\scp{\cdot}{\psi}\|^2 u_{\sphere(\Hilbert_2)}} \biggl[ 1_A \biggl( \frac{\scp{b_1}{\psi}}{\|\scp{b_1}{\psi}\|} \biggr) \biggr]\,,
\end{align}
and it was stated in \eqref{Psi2def}--\eqref{GAP3Psi} and proven earlier in this section that this quantity equals $\GAP{\rho_1^\psi}(A)$. 
\endproof

\section{Results}
\label{sec:results}

\subsection{GAP Measure From a Typical Wave Function of a Large System, Given the Reduced Density Matrix}
\label{sec:typical}

Let $\Hilbert_\total=\Hilbert_1\otimes\Hilbert_2$, where
$\Hilbert_1$ and $\Hilbert_2$ have respective dimension $d_1$
and $d_2$, with $d_1\leq d_2<\infty$. For any given density matrix $\rho_1$ on
$\Hilbert_1$, let
\begin{equation}
\mathscr{R}(\rho_1)=\bigl\{\psi\in\sphere(\Hilbert_\total):
\rho_1^\psi=\rho_1\bigr\}
\end{equation}
be the set of all normalized wave functions in $\Hilbert_{\total}$ with reduced density matrix $\rho_1^\psi=\rho_1$. We will see that $\mathscr{R}(\rho_1)$ is always non-empty.

Theorem~\ref{thm1} below concerns typical wave functions in
$\mathscr{R}(\rho_1)$, i.e., typical wave functions with fixed
reduced density matrix. The concept of ``typical'' refers to the 
uniform distribution $u_{\rho_1}$ on $\mathscr{R}(\rho_1)$;
an explicit definition of this distribution will be given in Section~\ref{sec:urho1def}.

Before we formulate Theorem~\ref{thm1}, we introduce some notation. First, for any Hilbert space $\Hilbert$, let $\D(\Hilbert)$ denote the set of all density operators on $\Hilbert$,
i.e., of all positive operators on $\Hilbert$ with trace 1. Second, when $\mu$ is a measure on $\Hilbert$ or $\sphere(\Hilbert)$ and $f(\psi)$ is a measurable function on $\Hilbert$ or $\sphere(\Hilbert)$ then we use the notation
\begin{equation}\label{muf}
\mu(f):=\int \mu(d\psi) f(\psi)\,.
\end{equation}
Third, let $\|f\|_\infty=\sup_{x} |f(x)|$.

\begin{thm}\label{thm1}
For every $\varepsilon>0$, all Hilbert spaces $\Hilbert_1,\Hilbert_2$ with dimensions \y{$1\leq d_1\leq d_2<\infty$ with $d_2\geq 4$}, every orthonormal basis $b=\{b_1,\ldots,b_{d_2}\}$ of $\Hilbert_2$, every $\rho_1\in \D(\Hilbert_1)$, and every bounded measurable test function $f:\sphere(\Hilbert_1)\to\RRR$,
\be\label{ineqthm1}
u_{\rho_1} \Bigl\{ \psi\in \mathscr{R}(\rho_1): \bigl|\mu_1^{\psi,b}(f) - \GAP{\rho_1}(f)\bigr|< \varepsilon \, \|f\|_\infty \Bigr\} \geq 1-\frac{4}{\varepsilon^2 d_2}\,.
\ee
In particular, for sufficiently big $d_2$ (uniformly in $b$ and $\rho_1$), the measure is arbitrarily close to 1.
\end{thm}

We give the proof, as well as those of Theorems~\ref{corbasis}--\ref{thm4}, in Section~\ref{sec:proofs}. 

It follows from Theorem~\ref{thm1} that, for every sequence $(\Hilbert_{2,n})_{n\in\NNN}$ of Hilbert spaces with $d_{2,n}=\dim \Hilbert_{2,n} \to \infty$ as $n\to\infty$ and every sequence $(b_n)_{n\in\NNN}$ of orthonormal bases $b_n=\{b_{1,n},\ldots,b_{d_{2,n},n}\}$ of $\Hilbert_{2,n}$, for every $\rho_1\in\D(\Hilbert_1)$, and for every bounded measurable function $f:\sphere(\Hilbert_1)\to\RRR$, the sequence of random variables $\mu_1^{\Psi_n,b_n}(f)$, where $\Psi_n$ has distribution $u_{\rho_1}$ on $\sphere(\Hilbert_1\otimes\Hilbert_{2,n})$, converges in distribution, as $n\to\infty$, to the constant $\GAP{\rho_1}(f)$, in fact uniformly in $\rho_1$, $b_n$ and those $f$ with $\|f\|_\infty\leq 1$. Because of the convergence for every $f$, we can say that the sequence of random measures $\mu_1^{\Psi_n,b_n}$ converges ``weakly in distribution'' to the fixed measure $GAP(\rho_1)$. 

A few comments about notation. In \cite{Gold1}, $d_1$ was called $k$,
$d_2$ was called $m$, and the notation for the
basis $\{b_1, \ldots, b_{d_2}\}$ was $\{\ket{1}, \ldots, \ket{m}\}$. For
enumerating the basis, we will use the letter $j$, and thus write
$b_j$; in \cite{Gold1}, the notation was $q_2$ for $j$ (subscript 2
because it refers to $\Hilbert_2$). For a random choice of $j$, we
write $J$; the corresponding notation in \cite{Gold1} was $Q_2$.

\subsection{GAP Measure From a Typical Basis of a Large System}

As already explained in \cite{Gold1}, instead of considering a
typical wave function and a fixed basis one can consider a fixed
wave function and a typical basis. Recall the notation $\rho_1^\psi=\tr_2 \ket{\psi}\bra{\psi}$.

\begin{thm}\label{corbasis}
For every $\varepsilon>0$, all Hilbert spaces $\Hilbert_1,\Hilbert_2$ of dimensions \y{$1\leq d_1\leq d_2<\infty$ with $d_2\geq 4$}, every $\psi\in \sphere(\Hilbert_1\otimes\Hilbert_2)$, and every bounded measurable test function $f:\sphere(\Hilbert_1)\to\RRR$,
\be\label{ineqcorbasis}
u_{ONB} \Bigl\{b \in ONB(\Hilbert_2): \bigl|\mu_1^{\psi,b}(f) - \GAP{\rho_1^\psi}(f)\bigr|< \varepsilon \, \|f\|_\infty \Bigr\} \geq 1-\frac{4}{\varepsilon^2 d_2}\,.
\ee
\end{thm}

\y{Our theorems are closely related to the phenomenon of \emph{concentration of measures} \cite{MS86}, which refers to the situation in certain metric probability spaces $X$ that an $\varepsilon$-neighborhood of a set of measure near 0 can have measure near 1 and leads to the consequence that relevant functions on these spaces are nearly constant (i.e., they are near their mean at most points). While our theorems are not implied by standard results about concentration of measures (see the end of Section~\ref{sec:pfcorbasis} for more detail), they are similar in that they say that certain functions are nearly constant, such as the function $\psi\mapsto \mu_1^{\psi,b}(f)$ on $X=\mathscr{R}(\rho_1)$ or the function $b\mapsto \mu_1^{\psi,b}(f)$ on $X=ONB(\Hilbert_2)$.}

\subsection{GAP Measure From a Typical Basis and a Typical Wave Function in a Large Subspace}

In our main physical application, the reduced density matrix $\rho_1^\psi$ is not fixed, although---by a fact known as canonical typicality---most of the relevant $\psi$s have a reduced density matrix $\rho_1^\psi$ that is close to a certain fixed density matrix, for example to the canonical density matrix $\rho_\beta= (1/Z)e^{-\beta H}$. In this section, we present two further universality theorems that are appropriate for such situations, in which the relevant set of $\psi$s is a subspace of $\Hilbert_1\otimes\Hilbert_2$ that will be denoted $\Hilbert_R$.
 
The physical setting to have in mind is this. A system with Hilbert space $\Hilbert_1$ is entangled with a large system whose Hilbert space is $\Hilbert_2$. The Hamiltonian $H$ is thus defined on $\Hilbert_{\total}=\Hilbert_1\otimes\Hilbert_2$; suppose the total system is confined to a finite volume, so that $H$ has pure point spectrum. Let $[E,E+\delta E]$ be a narrow energy window, located at a suitable energy $E$ such as one corresponding to a more or less fixed energy per particle or per volume. Then the \emph{micro-canonical energy shell} is the spectral subspace of $H$ associated with this interval, i.e., the subspace spanned by the eigenvectors with eigenvalues between $E$ and $E+\delta E$, and this is our subspace $\Hilbert_R$. The \emph{micro-canonical density matrix} $\rho_R$ is the density matrix associated with $\Hilbert_R$, i.e., $1/\dim\Hilbert_R$ times the projection to $\Hilbert_R$. Canonical typicality then asserts that, \y{if the interaction between the two systems can be neglected as in \eqref{noint}, then,} for most wave functions in $\sphere(\Hilbert_R)$, the reduced density matrix is approximately $\rho_\beta$ for an appropriate value of $\beta$.

For \y{general $\Hilbert_R$ (and regardless of the interaction)}, canonical typicality means that for most $\psi\in\sphere(\Hilbert_R)$, the reduced density matrix $\rho_1^\psi$ is close to $\tr_2\rho_R$. The precise statement that we make use of is Theorem~1 of \cite{PSW05} or the ``main theorem'' of \cite{PSW06}, which asserts, in a somewhat specialized and simplified form that suffices for our purposes:

\begin{lemma}\label{lem:can}
Consider a Hilbert space $\Hilbert_1$ of dimension $d_1\in\NNN$, another Hilbert space $\Hilbert_2$ of dimension $d_2\in\NNN$ and a subspace $\Hilbert_R\subseteq \Hilbert_1 \otimes \Hilbert_2$ of dimension $d_R$. Let $\rho_R$ be $1/d_R$ times the projection to $\Hilbert_R$, and $u_R$ the uniform distribution on $\sphere(\Hilbert_R)$. Then for every $\eta>0$,
\begin{equation}
  u_R \left\{ \psi \in \sphere(\Hilbert_R): \Bigl\|\rho_1^\psi -
  \tr_2 \rho_R  \Bigr\|_{\tr} \geq \eta + \frac{d_1}{\sqrt{d_R}}
  \right\} \leq 4 \exp\Bigl(-\frac{d_R\eta^2}{18\pi^3}\Bigr)\,.
\end{equation}
\end{lemma}

Here, the \emph{trace norm} is defined by
\begin{equation}
  \|M\|_{\tr} = \tr |M| = \tr \sqrt{M^* M}\,.
\end{equation}
By the \emph{uniform distribution} $u_R$ we mean the $(2d_R-1)$-dimensional surface area measure on $\sphere(\Hilbert_R)$, normalized so that $u_R(\sphere(\Hilbert_R))=1$.

\begin{thm}\label{thm3}
For every $0<\varepsilon<1$, $0<\delta<1$, $d_1\in\NNN$, every Hilbert space $\Hilbert_1$ with $\dim \Hilbert_1=d_1$, and every continuous function $f:\sphere(\Hilbert_1)\to\RRR$,  there \y{is a number $D_R=D_R(\varepsilon,\delta,d_1,f)>0$ such that for every $d_R\in\NNN$ with $d_R>D_R$}
and for every $\Hilbert_2$ and $\Hilbert_R\subseteq \Hilbert_1\otimes\Hilbert_2$ with \y{$\dim\Hilbert_{R} = d_{R}$,} 
\begin{multline}\label{eq:thm3}
  u_R \times u_{ONB} \Bigl\{ \bigl( \psi, b \bigr) \in
  \sphere(\Hilbert_R) \times ONB (\Hilbert_2) : \\
  \bigl|\mu_1^{\psi,b}(f) - \GAP{\tr_2 \rho_R}(f) \bigr| < \varepsilon \Bigr\}
  \geq 1-\delta\,.
\end{multline}
\end{thm}

It follows that, for every sequence $(\Hilbert_{2,n})_{n\in\NNN}$ of Hilbert spaces with $d_{2,n}=\dim 
\Hilbert_{2,n} \to \infty$ as $n\to\infty$, every sequence $(\Hilbert_{R,n})_{n\in\NNN}$ of subspaces of $\Hilbert_1\otimes\Hilbert_{2,n}$ with $d_{R,n}=\dim\Hilbert_{R,n} \to\infty$ as $n\to\infty$, and every continuous function $f:\sphere(\Hilbert_1)\to\RRR$, the sequence of random variables
\be
\mu_1^{\Psi_n,B_n}(f)-\GAP{\tr_2 \rho_{R,n}}(f)\,,
\ee
where $(\Psi_n,B_n)$ has distribution $u_{R,n}\times u_{ONB,n}$ on $\sphere(\Hilbert_{R,n})\times ONB(\Hilbert_{2,n})$, converges to zero in distribution as $n\to\infty$. We say that the sequence of random signed measures $\mu_1^{\Psi_n,B_n}-\GAP{\tr_2 \rho_{R,n}}$ converges ``weakly in distribution'' to zero.

\bigskip

For $0<\gamma<1/\dim\Hilbert$ let $\D_{\geq \gamma}(\Hilbert)$ denote the set of density matrices $\rho\in\D(\Hilbert)$ whose eigenvalues are all greater than or equal to $\gamma$ (so that, in particular, zero is not an eigenvalue of $\rho$).

\begin{thm}\label{thm4}
For every $0<\varepsilon<1$, $0<\delta<1$, $d_1\in\NNN$, and $0<\gamma<1/d_1$, there are \y{numbers $D_R'=D_R'(\varepsilon,\delta,d_1,\gamma)>0$ and $r'=r'(\varepsilon,d_1,\gamma)>0$ such that for every $d_R\in\NNN$ with $d_R>D'_R$,} 
for every Hilbert space $\Hilbert_1$ with $\dim \Hilbert_1=d_1$, for every $\Omega \in \D_{\geq\gamma}(\Hilbert_1)$, for every $\Hilbert_2$ and $\Hilbert_R\subseteq \Hilbert_1\otimes\Hilbert_2$ with \y{$\dim\Hilbert_{R} = d_{R}$} satisfying
\be\label{ROmega}
\bigl\|\tr_2 (\rho_R)-\Omega\bigr\|_{\tr}<r'\,,
\ee 
and for every bounded measurable function $f:\sphere(\Hilbert_1)\to\RRR$, 
\begin{multline}\label{eq:thm4}
  u_R \times u_{ONB} \Bigl\{ \bigl( \psi, b \bigr) \in
  \sphere(\Hilbert_R) \times ONB (\Hilbert_2) : \\
  \bigl|\mu_1^{\psi,b}(f) - \GAP{\Omega}(f) \bigr| < \varepsilon \,\|f\|_\infty \Bigr\}
  \geq 1-\delta\,.
\end{multline}
\end{thm}

If we want to consider just one particular density matrix $\Omega$ (of which zero is not an eigenvalue) then we can set $\gamma$ equal to the smallest eigenvalue of $\Omega$. It then follows that, for every sequence $(\Hilbert_{2,n})_{n\in\NNN}$ of Hilbert spaces with $d_{2,n}=\dim \Hilbert_{2,n} \to \infty$ as $n\to\infty$, and every sequence $(\Hilbert_{R,n})_{n\in\NNN}$ of subspaces of $\Hilbert_1\otimes\Hilbert_{2,n}$ with $d_{R,n}=\dim\Hilbert_{R,n} \to\infty$ and $\tr_2 \rho_{R,n} \to \Omega$ as $n\to\infty$, the sequence of random measures $\mu_1^{\Psi_n,B_n}$ converges weakly in distribution to the fixed measure $GAP(\Omega)$. In short,
\begin{equation}
\mu_1^{\psi,b}\stackrel{u_R\times u_{ONB}}{\Longrightarrow} GAP(\Omega)\,.
\end{equation}

Of the two theorems above, Theorem~\ref{thm3} is the simpler and perhaps more natural mathematical statement: it does not even mention any other density matrix than $\tr_2\rho_R$; its structure is to ask first that $\varepsilon$, $\delta$, and $f$ be specified, which define the accuracy of the desired approximations;\footnote{How $f$ defines a sense of accuracy becomes manifest if we consider finitely many test functions $f_1,\ldots,f_\ell$, assume $d_R>\max(D_R(f_1),\ldots,D_R(f_\ell))$, and then apply Theorem~\ref{thm3} to obtain that $\mu_1^{\psi,b}$ and $GAP(\tr_2\rho_R)$ agree approximately on all linear combinations of $f_1,\ldots,f_\ell$.} and it applies to all subspaces $\Hilbert_R$ of sufficient dimension. For the physical application, though, we often want to compare $\mu_1^\psi$ to $GAP(\Omega)$ rather than $GAP(\tr_2\rho_R)$, for example because $\Omega$ is the thermal density matrix $\rho_\beta=(1/Z)e^{-\beta H}$ while $\tr_2\rho_R$ is something complicated; we usually do not need that the estimate applies uniformly to all spaces $\Hilbert_R$ of sufficient dimension, but instead consider only one fixed $\Hilbert_R$; and in that situation we can, in fact, obtain an estimate, the one provided by Theorem~\ref{thm4}, that is uniform in $f$.

\subsection{GAP Measure as the Thermal Equilibrium Distribution}
\label{sec:thermo}

Theorem~\ref{thm4} justifies regarding $GAP(\rho_\beta)$ as the thermal
equilibrium distribution of the wave function of system 1 in the following way.
Let $\Hilbert_R$ be the microcanonical subspace, i.e., the spectral
subspace of $H$ associated with the interval $[E,E +\delta E]$. It
is a standard fact (e.g., \cite{G95,ML79}) that when the interaction energy between system 1 and system 2 is sufficiently small, i.e., when we may set
\be
H = H_1 \otimes I_2+ I_1\otimes H_2
\ee
on $\Hilbert_\total=\Hilbert_1\otimes\Hilbert_2$,  and 
when the eigenvalues of $H_2$ are sufficiently dense, then $\tr_2 \rho_R$ is approximately of the exponential form $Z^{-1}\exp(-\beta
H_1)$ with $Z = \tr \exp(-\beta H_1)$ for suitable $\beta>0$, i.e., is approximately the
canonical density matrix $\rho_\beta$. Then by Theorem~\ref{thm4} in
this special case of negligible interaction we have that for most wave functions $\psi\in\sphere(\Hilbert_R)$,
\begin{equation}
\mu_1^{\psi,b}\approx GAP(\rho_\beta)
\end{equation}
for most orthonormal bases $b$ of $\Hilbert_2$.

\section{Proofs}
\label{sec:proofs}

\subsection{Definition of $u_{\rho_1}$}
\label{sec:urho1def}

According to the Schmidt decomposition \cite{Schmidt}, every $\psi\in\Hilbert_\total$ can
be written in the form
\begin{equation}\label{Schmidt}
\psi = \sum_{i=1}^{d_1} c_i \, \tilde{\chi}_i\otimes\tilde{\phi}_i
\end{equation}
where $\{\tilde{\chi}_i\}$ is an orthonormal basis in $\Hilbert_1$,
$\{\tilde{\phi}_i\}$ is an orthonormal system in $\Hilbert_2$ (i.e., a set of
orthonormal vectors that is not necessarily complete), and the $c_i$ are
coefficients which can be chosen to be real and non-negative. If $\|\psi\|=1$, the reduced
density matrix of the system 1 is then
\begin{equation}
\rho_1^\psi=\sum_{i=1}^{d_1} c_i^2 \ket{\tilde{\chi}_i}\bra{\tilde{\chi}_i}\,.
\end{equation}
Thus, $\{\tilde{\chi}_i\}$ is an eigenbasis of $\rho_1^\psi$, and $c_i^2$ are the corresponding eigenvalues. 

Now let a density matrix $\rho_1$ be given, let $\{\chi_i\}$ be an eigenbasis for $\rho_1$, and let $0\leq p_i\leq 1$ be the corresponding eigenvalues. Then every $\psi\in\mathscr{R}(\rho_1)$ possesses a Schmidt decomposition of the form
\be\label{Schmidt2}
\psi=\sum_{i=1}^{d_1} \sqrt{p_i} \, \chi_i \otimes \phi_i
\ee
with some orthonormal system $\{\phi_i\}$ in $\Hilbert_2$. Indeed, we know it has a Schmidt decomposition \eqref{Schmidt} in which $\{\tilde{\chi}_i\}$ is an eigenbasis of $\rho_1$, and $c_i^2$ are the eigenvalues. Reordering the terms in \eqref{Schmidt}, we can make sure that $c_i=\sqrt{p_i}$. Any two eigenbases $\{\chi_i\}$ and $\{\tilde{\chi}_i\}$ of $\rho_1$ are related by a block unitary; more precisely, for every eigenvalue $p$ of $\rho_1$, $\{\chi_i:i\in\mathscr{I}(p)\}$ and $\{\tilde{\chi}_i:i\in\mathscr{I}(p)\}$ (using the index set $\mathscr{I}(p)=\{i:c_i^2=p\}=\{i:p_i=p\}$) are two orthonormal bases of the eigenspace of $p$, and thus related by a unitary matrix $(U_{ij}^{(p)})_{i,j\in\mathscr{I}(p)}$:
\be\label{chitildechi}
\tilde{\chi}_i = \sum_{j\in\mathscr{I}(p)} U^{(p)}_{ij} \, \chi_j\,.
\ee
Setting
\be\label{phitildephi}
\phi_i = \sum_{j\in\mathscr{I}(p)} U^{(p)}_{ji} \tilde{\phi}_j\,,
\ee
we obtain \eqref{Schmidt2}, and that $\{\phi_i\}$ is an orthonormal system.

Conversely, every orthonormal system $\{\phi_i\}$ in $\Hilbert_2$ defines, by \eqref{Schmidt2}, a $\psi\in\mathscr{R}(\rho_1)$. Thus, \eqref{Schmidt2} defines a bijection $F_{\rho_1,\{\chi_i\}}:ONS(\Hilbert_2,d_1)\to \mathscr{R}(\rho_1)$. The Haar measure on the unitary group of $\Hilbert_2$ defines the uniform distribution on the set of orthonormal bases of $\Hilbert_2$, of which the uniform distribution on $ONS(\Hilbert_2,d_1)$ is a marginal; let $u_{\rho_1,\{\chi_i\}}$ be its image under $F_{\rho_1,\{\chi_i\}}$. 

We note that $u_{\rho_1,\{\chi_i\}}$ actually does not depend on the choice of the eigenbasis $\{\chi_i\}$. Indeed, if $\{\tilde{\chi}_i\}$ is any other eigenbasis of $\rho_1$ (without loss of generality numbered in such a way that the eigenvalue of $\tilde{\chi}_i$ is $p_i$) then, as explained above, it is related to $\{\chi_i\}$ by a block unitary $d_1\times d_1$ matrix $U$ consisting of the blocks $(U_{ij}^{(p)})$. Let $\overline{U}$ be the matrix whose entries are the complex conjugates of the entries of $U$, and let $\hat{\overline{U}}$ denote the action of $\overline{U}$ on $ONS(\Hilbert_2,d_1)$ given by
\be
\hat{\overline{U}}\Bigl(\bigl\{\phi_i:i=1,\ldots,d_1\bigr\}\Bigr) = 
\biggl\{\sum_{j=1}^{d_1}\overline{U}_{ij} \phi_j:i=1,\ldots,d_1\biggr\}\,.
\ee
Then
\be
F_{\rho_1,\{\chi_i\}}=F_{\rho_1,\{\tilde{\chi}_i\}}\circ \hat{\overline{U}}\,.
\ee
Since the Haar measure is invariant under left multiplication, its marginal on $ONS(\Hilbert_2,d_1)$ is invariant under $\hat{\overline{U}}$. We thus define $u_{\rho_1}$ to be $u_{\rho_1,\{\chi_i\}}$ for any eigenbasis $\{\chi_i\}$.

\subsection{Proof of Theorem~\ref{corbasis}}
\label{sec:pfcorbasis}

\y{We first prove Theorem~\ref{corbasis} and later show that Theorem~\ref{thm1} is equivalent.} Let $\Var(Y)$ denote the variance of the random variable $Y$ and $\Cov(X,Y)$ the covariance of the random variables $X,Y$.

\begin{lemma}\label{lem:umoments}
Let $\Psi^u=(\Psi_1^u,\ldots,\Psi_d^u)\sim u_{\sphere(\CCC^d)}$. Then $\EEE\, \Psi^u_1=0$, $\EEE\, |\Psi^u_1|^2 = 1/d$,
\be\label{VarPsi12}
\EEE|\Psi^u_1|^4=\frac{2}{d(d+1)}\,,\quad
\Var\bigl(|\Psi^u_1|^2\bigr) = \frac{1}{d^2} \frac{d-1}{d+1}\,,
\ee
and
\be\label{CovPsi12Psi22}
\EEE\Bigl[ |\Psi^u_1|^2 |\Psi^u_2|^2 \Bigr] = \frac{1}{d(d+1)}\,,\quad
\Cov \Bigl(|\Psi^u_1|^2, |\Psi^u_2|^2 \Bigr) = -\frac{1}{d^2(d+1)}\,.
\ee 
\end{lemma}

\begin{proof}
\y{Since these relations can be found in many sources, e.g., \cite[Eq.~(2.3.6)]{GLTZ14} or \cite{vN29} (see Eq.~(144) and (149) in the English translation with $s=1$), we only give a brief outline.}
The relation $\EEE \, \Psi^u_1=0$ follows from the spherical symmetry of the distribution, and $\EEE\, |\Psi^u_1|^2=1/d$ from $\EEE\, \sum_{k=1}^d |\Psi^u_k|^2 =1$ and the fact that the $\Psi^u_k$ are exchangeable. \y{The first equation in \eqref{VarPsi12} can be obtained by means of integration in spherical coordinates in $\RRR^{2d}$, the second equation follows from the first. The first equation in \eqref{CovPsi12Psi22} follows easily from \eqref{VarPsi12} using that $\sum_{k=1}^d |\Psi_k^u|^2=1$ and thus $\EEE[(\sum |\Psi_k^u|^2)^2]=1$, and the second again from the first.}
\end{proof}

As a remark on Lemma~\ref{lem:umoments}, readers may find it useful to compare these results to the well-known fact that for large $d$, $\Psi^u_1$ and $\Psi^u_2$ are approximately distributed like independent complex Gaussian random variables $G_1,G_2$ with mean 0 and variance $\EEE\, |G_i|^2=1/d$. The  relations for $G_1,G_2$ corresponding to \eqref{VarPsi12}--\eqref{CovPsi12Psi22} are
\be\label{VarCovG}
\Var\bigl(|G_1|^2\bigr)=\frac{1}{d^2}\text{ and }\Cov\bigl(|G_1|^2,|G_2|^2\bigr)=0\,. 
\ee
Eq.~\eqref{CovPsi12Psi22} implies that the correlation coefficient of $|\Psi^u_1|^2$ and $|\Psi^u_2|^2$ is small like $1/d$, in agreement with the statement that they are approximately independent.

\y{Our proof of Theorem~\ref{corbasis} is based on the following lemma, which was proved in \cite{GLTZ14} as Theorem~1 (Version~3).

\begin{lemma}\label{lem:randombasis}
Let $\varepsilon>0,\delta>0$, $d\in\NNN$ with $d\geq 4$ and $d\geq 2\delta^{-2}\varepsilon^{-1}$, and let $\{b_1,\ldots,b_d\}$ be a random, uniformly distributed orthonormal basis of $\CCC^d$. Then, for every test function $\varphi\in L^2(\sphere(\CCC^d),u,\RRR)$, 
\be\label{thm1eqvarphi}
\PPP\biggl( \Bigl| \frac{1}{d} \sum_{j=1}^d \varphi(b_j) - 
\EEE_u(\varphi)\Bigr|\leq \delta \sqrt{\Var_u(\varphi)}\biggr) \geq1-\varepsilon\,,
\ee
where $\EEE_u(\varphi)$ and $\Var_u(\varphi)$ mean the mean and variance, respectively, relative to the uniform probability distribution over the unit sphere in $\CCC^d$.
\end{lemma}
}

\proof[Proof of Theorem~\ref{corbasis}]
\y{Fix $\varepsilon$, $\Hilbert_1$, $\Hilbert_2$, $\psi$, and $f$. Let the function $\varphi$ be defined, for any $\phi\in\sphere(\Hilbert_2)$, by
\be
\varphi(\phi) = d_2\; \bigl\| \scp{\phi}{\psi}\bigr\|^2 \, f\Bigl(P(\scp{\phi}{\psi})\Bigr)
\ee
with $P(\Psi)=\Psi/\|\Psi\|$ the projection to the unit sphere. 
Then, for any $b\in ONB(\Hilbert_2)$,
\be\label{mu1varphi}
\mu_1^{\psi,b}(f) = \frac{1}{d_2} \sum_{j=1}^{d_2} \varphi(b_j)\,,
\ee
cf.~\eqref{mu1expr}. Now regard $b$ as random, $b\sim u_{ONB}$. By Lemma~\ref{lem:EmuGAP},
\be\label{GAPEuvarphi}
GAP(\rho_1^\psi)(f)=\EEE\mu_1^{\psi,b}(f)=\EEE \frac{1}{d_2} \sum_{j=1}^{d_2} \varphi(b_j) = \EEE\varphi(b_1) = \EEE_u(\varphi)\,.
\ee

We now show that
\be\label{Varbound}
\Var_u(\varphi) \leq 2\, \|f\|_\infty^2\,.
\ee
Indeed, writing $X$ for a uniformly distributed random point on $\sphere(\Hilbert_2)$, and $Y=\scp{X}{\psi}_2 \in\Hilbert_1$, we have that
\begin{align}
\Var_u(\varphi) 
&=\EEE\bigl[ \varphi(X)^2\bigr]-\Bigl(\EEE[\varphi(X)] \Bigr)^2\\
&\leq \EEE\bigl[ \varphi(X)^2\bigr]\\
&= \EEE\Bigl[ d_2^2\, \| Y\|^4 \; f\bigl(P(Y) \bigr)^{\!\! 2}\Bigr]\\
&\leq d_2^2\; \|f\|_\infty^2 \; \EEE\Bigl[ \| Y\|^4 \Bigr]\,.
\label{lastlinevarphi}
\end{align}
We now estimate $\EEE \|Y\|^4$. As a tool, let
\be
\psi = \sum_{i=1}^{d_1} \sqrt{p_i} \, \chi_i \otimes \phi_i
\ee
be the Schmidt decomposition \cite{Schmidt} of $\psi$, where $(\chi_1,\ldots,\chi_{d_1})\in ONB(\Hilbert_1)$, $(\phi_1,\ldots,\phi_{d_2})\in ONB(\Hilbert_2)$, and $\rho_1^\psi=\sum_i p_i |\chi_i\rangle\langle \chi_i|$. Note that $\sum_{i=1}^{d_1} p_i =1$. Let
\be
p^2:= \sum_{i=1}^{d_1} p_i^2
\ee
and note that $0<p^2\leq 1$. Then
\begin{align}
\EEE \, \|Y\|^4
&=\EEE\,\bigl\| \scp{X}{\psi} \bigr\|^4 \\
&= \EEE\biggl[\Bigl(\sum_{i=1}^{d_1} p_i\,\bigl |\scp{X}{\phi_i}\bigr|^2\Bigr)^2\biggr]\\
&= \EEE\biggl[\sum_{i,j=1}^{d_1} p_ip_j\,\bigl |\scp{X}{\phi_i}\bigr|^2\bigl |\scp{X}{\phi_j}\bigr|^2\biggr]\\
&= \sum_{i=1}^{d_1} p_i^2 \,\EEE \bigl |\scp{X}{\phi_i}\bigr|^4 + \sum_{\substack{i,j=1\\i\neq j}}^{d_1} p_ip_j\,\EEE\biggl[\bigl |\scp{X}{\phi_i}\bigr|^2\bigl |\scp{X}{\phi_j}\bigr|^2\biggr]\\
&= \sum_{i=1}^{d_1} p_i^2 \,\EEE \bigl |\scp{X}{\phi_1}\bigr|^4 + \sum_{\substack{i,j=1\\i\neq j}}^{d_1} p_ip_j\,\EEE\biggl[\bigl |\scp{X}{\phi_1}\bigr|^2\bigl |\scp{X}{\phi_2}\bigr|^2\biggr]\\
\intertext{[because the distribution of $X$ is invariant under unitaries]}
&= p^2\, \EEE \bigl |\scp{X}{\phi_1}\bigr|^4 + (1-p^2)\,\EEE\biggl[\bigl |\scp{X}{\phi_1}\bigr|^2\bigl |\scp{X}{\phi_2}\bigr|^2\biggr]\\
&= p^2\, \EEE |X_1|^4 + (1-p^2) \,\EEE \Bigl[ |X_1|^2 \, |X_2|^2\Bigr]\\
&=\frac{2p^2}{d_2(d_2+1)} + \frac{1-p^2}{d_2(d_2+1)} \leq \frac{2}{d_2^2}
\label{lastlineEbpsi4}
\end{align}
by Lemma~\ref{lem:umoments}, with $X=(X_1,\ldots,X_{d_2})$.
This, together with \eqref{lastlinevarphi}, proves \eqref{Varbound}.

Now, in Lemma~\ref{lem:randombasis}, replace $d$ by $d_2$ (so $\CCC^d=\Hilbert_2$), replace $\varepsilon$ by $4\varepsilon^{-2} d_2^{-1}$, and $\delta$ by $\varepsilon/\sqrt{2}$. Then, the condition $d\geq 2\delta^{-2}\varepsilon^{-1}$ gets replaced by
\be
d_2\geq 2(\varepsilon/\sqrt{2})^{-2} (4\varepsilon^{-2} d_2^{-1})^{-1}\,,
\ee
which is satisfied because the right-hand side simplifies to $d_2$, and the condition $d\geq 4$ is satisfied as well. Inserting \eqref{mu1varphi} and \eqref{GAPEuvarphi}, Lemma~\ref{lem:randombasis} asserts that
\be
\PPP\biggl( \Bigl| \mu_1^{\psi,b}(f) - 
GAP(\rho_1^\psi)(f)\Bigr|\leq \frac{\varepsilon }{\sqrt{2}} \sqrt{\Var_u(\varphi)}\biggr) \geq1-\frac{4}{\varepsilon^2 d_2}\,.
\ee
From this and \eqref{Varbound}, we obtain \eqref{ineqcorbasis}, the relation we wanted to prove.}
\endproof

An alternative proof of Theorem~\ref{corbasis} is provided in an earlier version of this article that is available as a preprint at \url{http://arxiv.org/abs/1104.5482v1}. That proof did not make use of the theorem \cite{GLTZ14} about the uniformity of a random orthonormal basis quoted above as Lemma~\ref{lem:randombasis}, but instead of the theorem \cite{Colthesis} (see also the references in our preprint and in \cite{Colthesis}) that for a random $n\times n$ unitary matrix with
distribution given by the Haar measure on the unitary group $U(n)$,
the upper left (or any other) $k\times k$ submatrix, multiplied by a normalization factor $\sqrt{n}$, converges as $n\to\infty$ to a
matrix of independent complex Gaussian random variables with mean 0
and variance 1. (To understand the factor $\sqrt{n}$, note that a column of a unitary $n \times n$ matrix is a unit vector, and thus a single entry should be of order $1/\sqrt{n}$.) 

Another strategy for proving Theorem~\ref{corbasis} has been suggested by an anonymous referee and is based on \emph{concentration of measures} \cite{MS86}. The latter is a name for the fact that, in certain metric probability spaces $X$ including $X=\sphere(\CCC^d)$ and $X=ONB(\CCC^d)$ for large $d$ with the uniform measure $u_X$, the $\varepsilon$-neighborhood of any measurable subset $A\subseteq X$ of measure $u_X(A)\geq 1/2$ has measure close to 1. As a consequence, any 1-Lipschitz function $g$ (i.e., function with Lipschitz constant 1) on $X$ will be nearly constant, i.e., will stay within the $\varepsilon$-neighborhood of its median (or, for that matter, of its mean) on a set of measure close to 1. For our purposes, consider $X=ONB(\Hilbert_2)$ and $g(b) = \mu_1^{\psi,b}(f)$. Since the mean of $g$ is, by Lemma~\ref{lem:EmuGAP}, $GAP(\rho_1^\psi)(f)$, we would obtain that $u_{ONB}\{b: \mu_1^{\psi,b}-GAP(\rho_1^\psi)(f) \text{ is small}\}$ is close to 1, \y{provided that $g$ is 1-Lipschitz. But} $g$ will not be 1-Lipschitz unless $f$ is, so this argument requires a much stronger hypothesis on $f$ than Theorem~\ref{corbasis}. In fact, to have the statement of Theorem~\ref{corbasis} only for 1-Lipschitz test functions $f$ is rather useless because when (say) $\dim\Hilbert_1>10^5$ (as would realistically be the case in many applications of interest) then, by concentration of measures again, such an $f$ is nearly constant on $\sphere(\Hilbert_1)$ and thus unable to detect the difference between two measures such as $\mu_1^{\psi,b}$ and $GAP(\rho_1^\psi)$; that is, for a 1-Lipschitz function $f$, $\mu_1^{\psi,b}(f)-GAP(\rho_1^\psi)(f)$ \y{may be expected to} be small even if $\mu_1^{\psi,b}$ and $GAP(\rho_1^\psi)$ are not close to each other. That is why we follow a different strategy \y{and obtain Theorem~\ref{corbasis}, a stronger and more relevant result.}

\subsection{Proof of Theorem~\ref{thm1}}

\proof 
Note that for any unitary $U$ on $\Hilbert_2$
\begin{equation}\label{unitaryequiv}
\scp{U^{-1}b_j}{\psi} = \scp{b_j}{\,I_1\otimes U\,\psi}\, .
\end{equation}
From this fact and the fact that the Haar measure is invariant under $U\mapsto
U^{-1}$ it follows that the distribution of $\mu_1^{\psi,b}$, when $\psi\in\mathscr{R}(\rho_1)$
is $u_{\rho_1}$-distributed and $b$ is fixed, is the same as when $b$ is $u_{ONB}$-distributed and $\psi\in\mathscr{R}(\rho_1)$ is fixed. Thus, Theorem~\ref{corbasis} is equivalent to Theorem~\ref{thm1}. (It also follows that the distribution of $\mu_1^{\psi,b}$, when $\psi\in\mathscr{R}(\rho_1)$ is $u_{\rho_1}$-distributed and $b$ is fixed, does not depend on $b$, \y{and that its distribution, when $b$ is $u_{ONB}$-distributed and $\psi$ is fixed, does not depend on $\psi\in\mathscr{R}(\rho_1)$}.)
\endproof

\subsection{Continuity of GAP}

For the proofs of Theorems~\ref{thm3} and \ref{thm4}, we will exploit canonical typicality, i.e., the fact that for most $\psi\in\sphere(\Hilbert_R)$, the reduced density matrix $\rho_1^\psi$ is close to $\tr_2\rho_R$. Theorems~\ref{thm3} and \ref{thm4} then follow from Theorem~\ref{corbasis} via suitable continuity of the mapping $\rho\mapsto GAP(\rho)$. The following two lemmas provide somewhat different statements about continuity:  Recall that $\D_{\geq \gamma}(\Hilbert)$ is the set of density matrices with all eigenvalues greater than or equal to $\gamma$. Lemma~\ref{lem:contLinfty} asserts that $GAP(\rho)(f)$ depends in a \emph{uniformly} continuous way on both $\rho$ and $f$ when we restrict $\rho$ to $\D_\gamma(\Hilbert)$ for arbitrarily small $\gamma>0$; continuity is not uniform without this restriction. However, Lemma~\ref{lem:cont} asserts that for any \emph{fixed} and \emph{continuous} test function $f$, continuity is uniform in $\rho$ without restrictions.

\begin{lemma}\label{lem:cont}
For every $0<\varepsilon<1$, every $d\in\NNN$, every Hilbert space $\Hilbert$ with $\dim \Hilbert=d$, and every continuous function $f:\sphere(\Hilbert)\to\RRR$ there is $r=r(\varepsilon,d,f)>0$ such that for all $\rho,\Omega\in\D(\Hilbert)$,
\be
\text{if }\|\rho-\Omega\|_{\tr} < r
\text{ then }
\bigl| GAP(\rho)(f)-GAP(\Omega)(f) \bigr| < \varepsilon\,.
\ee
\end{lemma}

While all norms on $\D(\Hilbert)$ are equivalent for $\dim \Hilbert< \infty$, we use the trace norm $\|\cdot\|_{\tr}$ here because in this norm the continuity extends to $\dim \Hilbert=\infty$ and because it is used in Lemma~\ref{lem:can}.

To formulate the other continuity statement, let $u_{\sphere(\Hilbert)}$ denote the normalized uniform measure on the unit sphere in $\Hilbert$. For any density matrix $\rho\in\D(\Hilbert)$ of which zero is not an eigenvalue, $GAP(\rho)$ possesses a density relative to $u_{\sphere(\Hilbert)}$ \cite{Gold1}.

\begin{lemma}\label{lem:contLinfty}
For every $0<\varepsilon<1$, every $d\in\NNN$, every Hilbert space $\Hilbert$ with $\dim \Hilbert=d$, and every $0<\gamma<1/d$, there is $r=r(\varepsilon,d,\gamma)>0$ such that for all $\rho,\Omega\in\D_{\geq\gamma}(\Hilbert)$,  
\be\label{eq:contLinfty}
\text{if }\|\rho-\Omega\|_{\tr}<r
\text{ then }
\biggl\| \frac{dGAP(\rho)}{du_{\sphere(\Hilbert)}}
-\frac{dGAP(\Omega)}{du_{\sphere(\Hilbert)}} 
\biggr\|_{\infty} < \varepsilon\,.
\ee
As a consequence, for such $\rho$ and $\Omega$,
\be\label{eq:contL1}
\bigl| GAP(\rho)(f)-GAP(\Omega)(f) \bigr|<\varepsilon \, \|f\|_1
\ee
for every $f\in L^1\bigl(\sphere(\Hilbert),u_{\sphere(\Hilbert)}\bigr)$.
\end{lemma}

It follows in particular that for any \emph{fixed} density matrix $\Omega$ of which zero is not an eigenvalue and any sequence $(\rho_n)$ of density matrices with $\rho_n\to\Omega$, the density of $GAP(\rho_n)$ converges to that of $GAP(\Omega)$ in the $\|\cdot\|_\infty$ norm: Take $\gamma>0$ to be less than the smallest eigenvalue of $\Omega$ and note that only finitely many $\rho_n$ can lie outside $\D_{\geq \gamma}(\Hilbert)$.

To see that in Lemma~\ref{lem:contLinfty} $\D_{\geq\gamma}(\Hilbert)$ cannot be replaced by $\D(\Hilbert)$ (i.e., that continuity is not uniform without restrictions), note that, when 0 is an eigenvalue of $\Omega$, $\GAP{\Omega}$ does not have a density with respect to $u_{\sphere(\Hilbert)}$, so that at such an $\Omega$, $\rho\mapsto\GAP{\rho}$ is certainly not continuous in $L^\infty\bigl(\sphere(\Hilbert),u_{\sphere(\Hilbert)}\bigr)$, \y{nor in $L^1$ (which would correspond to the variation distance of measures).}

To see that in Lemma~\ref{lem:cont} one cannot drop the assumption that $f$ is continuous, consider an $\Omega$ that has zero as an eigenvalue and a $\rho$ that does not. Then $GAP(\Omega)$ is concentrated on a subspace of dimension less than $d$ while $GAP(\rho)$ has a density on the sphere and lies \emph{near} (rather than \emph{in}) that subspace. Thus, for a test function $f$ that is bounded measurable but not continuous, $GAP(\rho)(f)$ does not have to be close to $GAP(\Omega)(f)$.

\label{sec:contproof}

As part of the proof of Lemma~\ref{lem:cont}, we will need the continuity property of Gaussian measures expressed in the next lemma. When $\mu_n,\mu$ are measures on a topological space $X$, we write $\mu_n \Rightarrow \mu$ to denote that the sequence of measures $\mu_n$ converges weakly to $\mu$. This means that $\mu_n(f) \to \mu(f)$ for every bounded continuous function $f:X\to \RRR$ and implies that the same thing is true for every bounded measurable
function $f:X\to\RRR$ such that $\mu(D(f))=0$, where $D(f)$ is the set of
discontinuities of $f$.

\begin{lemma}\label{contGk}
The mapping $\rho\mapsto\G{\rho}$ is continuous in the weak topology on measures: If $\rho_n \in \D(\CCC^d)$ for every $n\in\NNN$ and $\rho_n \to \rho$ then $G(\rho_n) \Rightarrow G(\rho)$.
\end{lemma}

\proof 
We use characteristic functions; as usual, the characteristic
function $\hat{\mu}:\RRR^{2d} \to \CCC$ of a probability measure
$\mu$ on $\RRR^{2d}$ is defined by
\begin{equation}
  \hat{\mu}(k_1,\ldots, k_{2d}) = \int \mu(dx_1 \cdots dx_{2d}) \,
  \exp\Bigl(i\sum_{j=1}^{2d} k_j x_j \Bigr)\,,
\end{equation}
or, in our notation on $\Hilbert = \CCC^{d}$,
\begin{equation}
  \hat{\mu}(\phi) = \int \mu(d\psi) \,
  \exp\Bigl(i\Re \scp{\phi}{\psi} \Bigr)\,,
\end{equation}
where $\Re$ denotes the real part. We write $\mu_n = G(\rho_n)$ and
$\mu=G(\rho)$; their characteristic functions are:
\begin{equation}\label{hatGauss}
  \hat{\mu}_n(\psi) = \exp\bigl(-\bra{\psi}\rho_n\ket{\psi}\bigr)\,, \quad
  \hat{\mu}(\psi) = \exp\bigl(-\bra{\psi}\rho\ket{\psi}\bigr)\,.
\end{equation}
If $\rho_n \to \rho$ then $\bra{\psi}\rho_n\ket{\psi} \to
\bra{\psi}\rho\ket{\psi}$ for every $\psi$ and thus $\hat{\mu}_n \to \hat{\mu}$ pointwise. Since (e.g., \cite{Bill}) pointwise convergence of
the characteristic functions is equivalent (in finite dimension) to
weak convergence of the associated measures, it follows that
$G(\rho_n) \Rightarrow G(\rho)$, which is what we wanted to show.
\endproof

\proof[Proof of Lemma~\ref{lem:cont}]
Since $\D(\Hilbert)$ is compact, uniform continuity follows from continuity. That is, it suffices to show that, assuming $\rho_n \in \D(\Hilbert)$ for every $n\in\NNN$,
\begin{equation}\label{GAPcont}
\text{if }\rho_n \to \rho
\text{ then }GAP(\rho_n) \Rightarrow GAP(\rho)\,.
\end{equation}
This follows from Lemma~\ref{contGk}, the continuity of the adjustment mapping $A$ defined in \eqref{Adef} in Section~\ref{sec:GAPdef}, and the continuity of the projection $P:\Hilbert\setminus\{0\}\to\sphere(\Hilbert)$. Our first step is to establish the continuity of $A$ on the set of probability measures $\mu$ on $\Hilbert$ such that $\int \mu(d\psi) \, \|\psi\|^2 = 1$: If, for every $n\in\NNN$, $\mu_n$ is a probability measure on the Borel $\sigma$-algebra of $\Hilbert$ such that $\int \mu_n(d\psi) \, \|\psi\|^2 = 1$, then
\be\label{Acont}
\text{if }\mu_n \Rightarrow \mu
\text{ and }\int \mu(d\psi) \,\|\psi\|^2 = 1
\text{ then }A\mu_n\Rightarrow A\mu\,.
\ee

Fix $\varepsilon>0$ and an arbitrary non-zero, bounded, continuous
function $f: \Hilbert \to \RRR$. As before, we use the notation $N(\psi) = \|\psi\|$. Since,
by hypothesis, $\mu(N^2) = 1$, there
exists $R>0$ so large that
\begin{equation}
  \int_{\{\psi\in \Hilbert: \|\psi\|<R\}} \mu(d\psi) \, \|\psi\|^2 > 1- \frac{\varepsilon}{6\|f\|_\infty}\,.
\end{equation}
Let the ``cut-off function''
$\chi_0: [0,\infty) \to [0,1]$ be any continuous function such that
$\chi_0(x)=1$ for $x\leq R$ and $\chi_0(x)=0$ for $x\geq 2R$; set $\chi(\psi)=\chi_0(\|\psi\|)$. Because $\chi  N^2$ and $f\chi N^2$ are bounded continuous functions, and because
$\mu_n \Rightarrow \mu$, we have that $\mu_n(\chi  N^2) \to \mu(\chi 
N^2)$ and $\mu_n(f\chi N^2) \to \mu(f\chi N^2)$; that is, there is an $n_1\in \NNN$ such that, for all $n>n_1$,
\begin{equation}
  \left| \mu_n(\chi  N^2) - \mu(\chi  N^2) \right| < \frac{\varepsilon}{3\|f\|_\infty} 
\end{equation}
and
\begin{equation}
  \left| \mu_n(f\chi  N^2) - \mu(f\chi  N^2) \right| < \frac{\varepsilon}{3}\,.
\end{equation}
Thus, for all $n>n_1$, we have that
\begin{align}
  &\left| A\mu_n(f) - A\mu(f) \right| 
  = \left| \mu_n(f N^2) - \mu(f N^2) \right| \\
  &\leq \left| \mu_n(f\chi  N^2) - \mu(f\chi  N^2) \right| +
  \left| \mu_n\bigl( f(1-\chi) N^2 \bigr) \right| +
  \left| \mu\bigl( f(1-\chi)N^2 \bigr) \right| \\
  &< \frac{\varepsilon}{3} + \|f\|_\infty \mu_n\bigl( (1-\chi)N^2 \bigr) 
  + \|f\|_\infty \mu\bigl( (1-\chi)N^2\bigr)\\
  &= \frac{\varepsilon}{3} + \|f\|_\infty \bigl(1-\mu_n(\chi N^2)\bigr)
  + \|f\|_\infty \bigl(1-\mu(\chi N^2)\bigr)\\
  &\leq \frac{\varepsilon}{3} + 2\|f\|_\infty \bigl(1-\mu(\chi N^2)\bigr) 
  + \|f\|_\infty \bigl|\mu_n(\chi N^2)-\mu(\chi N^2)\bigr|\\
  &\leq \frac{\varepsilon}{3}+\frac{\varepsilon}{3}+\frac{\varepsilon}{3}
  = \varepsilon\,.
\end{align}
This proves \eqref{Acont}.\footnote{We remark that the 
hypothesis $\int \mu(d\psi) \|\psi\|^2=1$ cannot
be dropped, that is, does not follow from $\int \mu_n(d\psi)
\|\psi\|^2=1$. An example is $\mu_n = (1-1/n) \delta_0 + (1/n)
\delta_{\psi_n}$, where $\delta_\phi$ means the Dirac delta measure
at $\phi$ and $\psi_n$ is any vector with $\|\psi_n\|^2 = n$; then
$\mu_n$ is a probability measure with $\int \mu_n(d\psi)
\|\psi\|^2=1$ but $\mu_n \Rightarrow \delta_0$, which has $\int
\delta_0(d\psi) \|\psi\|^2=0$.}

We are now ready to establish \eqref{GAPcont}. Suppose $\rho_n\to\rho$. We
have that $\GAP{\rho_n}=P_*A(\G{\rho_n})$ and that $\left(AG(\rho)\right)(0)=0$. Since
$\psi\mapsto P\psi$ is continuous for $\psi\neq 0$, \eqref{GAPcont}
follows from \eqref{Acont} and Lemma~\ref{contGk}. This completes the proof of Lemma~\ref{lem:cont}.
\endproof

\bigskip

\proof[Proof of Lemma~\ref{lem:contLinfty}]
We first note that, for any self-adjoint $d\times d$ matrix $A$ and $\psi\in\sphere(\CCC^d)$,
\be\label{trnormineq}
\Bigl| \scp{\psi}{A|\psi} \Bigr| \leq \|A\| \leq \|A\|_{\tr}\,.
\ee

For any density matrix $\rho\in\D(\Hilbert)$ of which zero is not an eigenvalue, the density of $GAP(\rho)$ relative to $u_{\sphere(\Hilbert)}$ is given by \eqref{mupowerlaw}.
Using this expression, we will now show that \eqref{eq:contLinfty} holds when $\rho$ is sufficiently close to $\Omega$. 
This follows from the facts (i)~that, on $\D_{\geq \gamma}(\Hilbert)$, the functions $\rho\mapsto1/\det\rho$ and $\rho\mapsto\rho^{-1}$ are uniformly continuous, (ii)~that
\be
\Bigl| \scp{\psi}{\rho^{-1}|\psi}-\scp{\psi}{\Omega^{-1}|\psi}\Bigr| \leq \|\rho^{-1}-\Omega^{-1}\|_{\tr}
\ee
for all $\psi\in\sphere(\Hilbert)$, (iii)~that the function $x\mapsto x^{-d-1}$ is uniformly continuous on the interval $[1,\infty)$, and (iv)~that $\scp{\psi}{\rho^{-1}|\psi}\geq 1$, $\scp{\psi}{\Omega^{-1}|\psi}\geq 1$. 
This establishes the existence of $r(\varepsilon,d,\gamma)>0$ as described in Lemma~\ref{lem:contLinfty}. 

Now \eqref{eq:contL1} follows from \eqref{eq:contLinfty} according to
\begin{align}
&\bigl| GAP(\rho)(f)-GAP(\Omega)(f) \bigr|\nonumber\\
&=\Biggl|\,\,\int\limits_{\sphere(\Hilbert)}du_{\sphere(\Hilbert)} \biggl( \frac{dGAP(\rho)}{du_{\sphere(\Hilbert)}}(\psi)
-\frac{dGAP(\Omega)}{du_{\sphere(\Hilbert)}}(\psi)\biggr) f(\psi) \Biggr|\\
&\leq\int\limits_{\sphere(\Hilbert)}du_{\sphere(\Hilbert)} \biggl| \frac{dGAP(\rho)}{du_{\sphere(\Hilbert)}}(\psi)
-\frac{dGAP(\Omega)}{du_{\sphere(\Hilbert)}}(\psi)\biggr| \, |f(\psi)|
<\varepsilon \, \|f\|_1\,.
\end{align}
\endproof

\subsection{Proof of Theorem~\ref{thm3}}

\proof[Proof of Theorem~\ref{thm3}.] 
Suppose we are given $0<\varepsilon<1$, $0<\delta<1$, $d_1\in\NNN$, a Hilbert space $\Hilbert_1$ of dimension $d_1$, and a continuous function $f:\sphere(\Hilbert_1)\to\RRR$. \y{Set
\be
D_R(\varepsilon,\delta,d_1,f) =\max\Biggl\{ 4d_1, \frac{32d_1\|f\|_\infty^2}{\varepsilon^2 \delta} ,\frac{4d_1^2}{r(\varepsilon/2,d_1,f)^2},\frac{72\pi^3 \log (8/\delta)}{r(\varepsilon/2,d_1,f)^2}\Biggr\}\,,
\ee
with} $r(\varepsilon,d,f)$ as provided by Lemma~\ref{lem:cont}. Now consider \y{any $d_R\in\NNN$ with $d_R>D_R$ and}
any $\Hilbert_2$ and $\Hilbert_R\subseteq \Hilbert_1\otimes\Hilbert_2$ with \y{$\dim\Hilbert_{R}=d_{R}$; it follows that
\be\label{d2lowerbound}
d_2=\dim \Hilbert_2 \geq d_R/d_1 > \frac{32\|f\|_\infty^2}{\varepsilon^2 \delta}\,.
\ee
Let} $M(f,\varepsilon)$ be the set mentioned in \eqref{eq:thm3}, 
\be
M(f,\varepsilon)=
\left\{ \bigl( \psi, b \bigr) \in
  \sphere(\Hilbert_R) \times ONB (\Hilbert_2) : \bigl|
  \mu_1^{\psi,b}(f) - \GAP{\tr_2 \rho_R}(f) \bigr| < \varepsilon \right\},
\ee
let
\be
M'(f,\varepsilon) =
\left\{\bigl( \psi, b \bigr) \in
  \sphere(\Hilbert_R) \times ONB (\Hilbert_2) : \bigl|
  \mu_1^{\psi,b}(f) - \GAP{\rho_1^\psi}(f) \bigr| < \varepsilon \right\}
\ee
and
\be
M''(\varepsilon) =
\Bigl\{\psi\in\sphere(\Hilbert_R): \|\rho_1^\psi -
\tr_2 \rho_R\|_{\tr}<\varepsilon \Bigr\}.
\ee
Then, by Lemma~\ref{lem:cont},
\be\label{MM'M''}
M(f,\varepsilon) \supseteq 
M'\Bigl(f,\frac{\varepsilon}{2}\Bigr) \cap 
\Bigl[M''\Bigl(r\bigl(\frac{\varepsilon}{2},d_1,f\bigr)\Bigr)\times ONB(\Hilbert_2)\Bigr]\,.
\ee

Theorem~\ref{corbasis} yields, using \y{\eqref{d2lowerbound},}
that for every $\psi\in\sphere(\Hilbert_R)$,
\be
u_{ONB} \Bigl\{ b \in ONB(\Hilbert_2): \bigl|\mu_1^{\psi,b}(f) - \GAP{\rho_1^\psi}(f)\bigr|< \frac{\varepsilon}{2} \Bigr\} \geq 1-\delta/2\,.
\ee
Thus, averaging over $\psi\in\sphere(\Hilbert_R)$ according to $u_R$,
\be\label{M'}
u_R\times u_{ONB} \Bigl( M'(f,\varepsilon/2) \Bigr) \geq 1 -\delta/2\,.
\ee

Lemma~\ref{lem:can} with $\eta = r/2$ for $r=r(\varepsilon/2,d_1,f)$ yields, using our assumption $d_R > 4d_1^2/r^2$, which implies that $d_1/\sqrt{d_R}\leq r/2$, that
\be
u_R(M''(r)) \geq 1 - 4\exp\Bigl(-\frac{d_R r^2}{18\pi^34}\Bigr)\,.
\ee
Using our assumption $d_R > 18\pi^34\log(8/\delta)/r^2$, the right-hand side is greater than or equal to $1-\delta/2$, and thus
\be\label{M''}
u_R\times u_{ONB} \Bigl[M''(r)\times ONB(\Hilbert_2)\Bigr] \geq 1- \delta/2\,.
\ee

From \eqref{M'}, \eqref{M''}, and \eqref{MM'M''} together we have that
\be
u_R\times u_{ONB} \Bigl[ M(f,\varepsilon) \Bigr] \geq 1 -\delta\,,
\ee
which is what we wanted to show.
\endproof

\subsection{Proof of Theorem~\ref{thm4}}

\proof[Proof of Theorem~\ref{thm4}]
Suppose we are given $0<\varepsilon<1$, $0<\delta<1$, $d_1\in\NNN$, $0<\gamma<1/d_1$, and a Hilbert space $\Hilbert_1$ of dimension $d_1$. \y{Set
\begin{align}
D'_R&=D'_R(\varepsilon,\delta,d_1,\gamma) = \max\Biggl\{ 4d_1, \frac{32d_1}{\varepsilon^2 \delta} ,\frac{4d_1^2}{(r')^2}, \frac{72\pi^3 \log (8/\delta)}{(r')^2} \Biggr\}\,,\\
r'&=r'(\varepsilon,d_1,\gamma)=\frac{1}{2}r(\varepsilon/2,d_1,\gamma)\,,
\end{align}
with} $r(\varepsilon,d,\gamma)$ as provided by Lemma~\ref{lem:contLinfty}. Now consider any \y{$d_R\in\NNN$ with $d_R>D'_R$,} 
any $\Omega\in\D_{\geq\gamma}(\Hilbert_1)$, any $\Hilbert_2$ and $\Hilbert_R\subseteq \Hilbert_1\otimes\Hilbert_2$ with \y{$\dim\Hilbert_{R}=d_{R}$,} and any bounded measurable function $f:\sphere(\Hilbert_1)\to\RRR$. \y{It follows that
\be\label{d2lower2}
d_2=\dim\Hilbert_2 \geq d_R/d_1 > \frac{32}{\varepsilon^2\delta}\,.
\ee
Let} $M_0(f,\varepsilon)$ be the set mentioned in \eqref{eq:thm4}, 
\be
M_0(f,\varepsilon)=
\left\{ \bigl( \psi, b \bigr) \in
  \sphere(\Hilbert_R) \times ONB (\Hilbert_2) : \bigl|
  \mu_1^{\psi,b}(f) - GAP(\Omega)(f) \bigr| < \varepsilon \, \|f\|_\infty \right\},
\ee
let, as in the proof of Theorem~\ref{thm3},
\be
M'(f,\varepsilon) =
\left\{\bigl( \psi, b \bigr) \in
  \sphere(\Hilbert_R) \times ONB (\Hilbert_2) : \bigl|
  \mu_1^{\psi,b}(f) - GAP(\rho_1^\psi)(f) \bigr| < \varepsilon \right\},
\ee
let
\be
M_0''(\varepsilon) =
\Bigl\{\psi\in\sphere(\Hilbert_R): \|\rho_1^\psi -
\Omega\|_{\tr}<\varepsilon \Bigr\},
\ee
and let, as in the proof of Theorem~\ref{thm3},
\be
M''(\varepsilon) =
\Bigl\{\psi\in\sphere(\Hilbert_R): \|\rho_1^\psi -
\tr_2 \rho_R\|_{\tr}<\varepsilon \Bigr\}.
\ee
Now assume $\bigl\|\tr_2 \rho_R-\Omega\bigr\|_{\tr}<r'$. Then
\be\label{M0''M''}
M_0''(2r') \supseteq M''(r')
\ee
and, by Lemma~\ref{lem:contLinfty} and $\|f\|_1\leq \|f\|_\infty$,
\be\label{M0M'M0''}
M_0(f,\varepsilon) \supseteq 
M'\biggl(f,\frac{\varepsilon\|f\|_\infty}{2}\biggr) \cap 
\Bigl[M_0''(2r')\times ONB(\Hilbert_2)\Bigr]\,.
\ee

As in the proof of Theorem~\ref{thm3}, Theorem~\ref{corbasis} yields \eqref{M'} with $\varepsilon$ replaced by $\varepsilon\|f\|_\infty$ using \y{\eqref{d2lower2},} 
and Lemma~\ref{lem:can} yields \eqref{M''} with $r$ replaced by $r'$, using our assumption $d_R>D'_R$. From \eqref{M'}, \eqref{M''}, \eqref{M0''M''}, and \eqref{M0M'M0''} together we have that
\be
u_R\times u_{ONB} \Bigl[ M_0(f,\varepsilon) \Bigr] \geq 1 -\delta\,,
\ee
which is what we wanted to show.
\endproof

\bigskip

\noindent\textit{Acknowledgments.} 
We are grateful to Beno\^\i t Collins and an anonymous referee for helpful remarks.
S.~Goldstein was supported in part by the National Science Foundation [grant DMS-0504504].
J.~L.~Lebowitz and C.~Mastrodonato were supported in part by the National Science Foundation [grant  DMR 08-02120] and the Air Force Office of Scientific Research [grant AF-FA 49620-01-0154].
N.~Zangh\`\i\ was supported in part by Istituto Nazionale di Fisica Nucleare.

\end{document}